\documentclass[12pt]{article}
\usepackage{times}
\usepackage{geometry}
\usepackage[utf8]{inputenc}
\usepackage{enumitem,amssymb}
\usepackage{ragged2e}
\newlist{thematic}{itemize}{8}
\setlist[thematic]{label=$\square$}
\usepackage{enumitem}
\setlist{nosep}
\usepackage{pifont}
\usepackage{deluxetable}
\usepackage{aas_macros}

\usepackage{setspace}
\usepackage[hang,flushmargin]{footmisc}

\usepackage{sectsty}

\sectionfont{\fontsize{15}{20}\selectfont}
\subsectionfont{\fontsize{14}{16}\selectfont}

\usepackage{natbib}
\usepackage{sidecap}
\usepackage{graphicx}
\usepackage{floatrow}

\begin{document}
\pagestyle{empty}
\raggedright
\huge
Astro2020 Science White Paper \linebreak
\vspace{-15pt}

Imaging Giant Protoplanets with the ELTs\linebreak
\normalsize

\noindent \textbf{Thematic Areas:} \hspace*{60pt} $\boxtimes$ Planetary Systems \hspace*{10pt} $\boxtimes$ Star and Planet Formation \hspace*{20pt}\linebreak
$\square$ Formation and Evolution of Compact Objects \hspace*{31pt} $\square$ Cosmology and Fundamental Physics \linebreak
  $\square$  Stars and Stellar Evolution \hspace*{1pt} $\square$ Resolved Stellar Populations and their Environments \hspace*{40pt} \linebreak
  $\square$    Galaxy Evolution   \hspace*{45pt} $\square$             Multi-Messenger Astronomy and Astrophysics \hspace*{65pt} \linebreak

\textbf{Principal Author:}

Name: Steph Sallum	
 \linebreak						
Institution: University of California, Santa Cruz 
 \linebreak
Email: ssallum@ucsc.edu
 \linebreak
Phone: (831) 459-4820
 \linebreak
 
\vspace{-10pt}

\textbf{Co-authors:}
  \linebreak
Vanessa Bailey (Jet Propulsion Laboratory, California Inst.~of Technology)\\
Rebecca Bernstein (Carnegie Observatories)\\
Alan Boss (Carnegie Institution)\\
Brendan Bowler (The University of Texas at Austin)\\
Laird Close (University of Arizona)\\
Thayne Currie (NASA-Ames Research Center)\\
Ruobing Dong (University of Victoria)\\
Catherine Espaillat (Boston University)\\
Michael P. Fitzgerald (University of California, Los Angeles)\\
Katherine B. Follette (Amherst College)\\
Jonathan Fortney (University of California, Santa Cruz)\\
Yasuhiro Hasegawa (Jet Propulsion Laboratory, California Inst.~of Technology)\\
Hannah Jang-Condell (University of Wyoming)\\
Nemanja Jovanovic (California Inst.~of Technology)\\
Stephen R. Kane (University of California, Riverside)\\
Quinn Konopacky (University of California, San Diego)\\
Michael Liu (University of Hawaii)\\
Julien Lozi (National Astronomical Observatory of Japan)\\
Jared Males (University of Arizona)\\
Dimitri Mawet (California Inst.~of Technology / Jet Propulsion Laboratory)\\
Benjamin Mazin (University of California, Santa Barbara)\\
Max Millar-Blanchaer (Jet Propulsion Laboratory, California Inst.~of Technology)\\
Ruth Murray-Clay (University of California, Santa Cruz)\\
Garreth Ruane (Jet Propulsion Laboratory, California Inst.~of Technology)\\
Andrew Skemer (University of California, Santa Cruz)\\
Motohide Tamura (Univ. of Tokyo / Astrobiology Center of NINS)\\
Gautam Vasisht (Jet Propulsion Laboratory, California Inst. of Technology)\\
Jason Wang (California Inst.~of Technology)\\
Ji Wang (The Ohio State University)\\

\textbf{Abstract:} 
\linebreak
We have now accumulated a wealth of observations of the planet-formation environment and of mature planetary systems. 
These data allow us to test and refine theories of gas-giant planet formation by placing constraints on the conditions and timescale of this process.
Yet a number of fundamental questions remain unanswered about how protoplanets accumulate material, their photospheric properties and compositions, and how they interact with protoplanetary disks.
While we have begun to detect protoplanet candidates during the last several years, we are presently only sensitive to the widest separation, highest mass / accretion rate cases.
Current observing facilities lack the angular resolution and inner working angle to probe the few-AU orbital separations where giant planet formation is thought to be most efficient.
They also lack the contrast to detect accretion rates that would form lower mass gas giants and ice giants.
Instruments and telescopes coming online over the next decade will provide high contrast in the inner giant-planet-forming regions around young stars, allowing us to build a protoplanet census and to characterize planet formation in detail for the first time.

\pagebreak
\pagestyle{plain}
\setcounter{page}{1}

\section{Observations of Giant Planet Formation and Open Questions}
\vspace{-8pt}

During the last few decades, observations of planetary systems and protoplanetary disks have furthered our understanding of gas-giant formation by enabling tests of planet formation theory.
Planet demographics discriminate between bottom-up (e.g. core-accretion) and top-down (e.g. gravitational instability) formation scenarios and inform theories of migration \citep[e.g.][]{1996Icar..124...62P,2005ApJ...621L..69R}.
Millimeter observations of protoplanetary disks constrain the timescale for disk dissipation (and thus planet formation) to $\sim10$ Myr \citep[e.g.][]{2014MNRAS.445.3315N}.
ALMA images reveal disk structures such as gaps and holes, spirals, and warps \citep[e.g.][]{2015ApJ...808L...3A}, which have also been observed in infrared imaging and polarimetry \citep[e.g.][]{2012ApJ...748L..22M}.
Some disks show different structure in the infrared (small grains) compared to the millimeter (large grains), and some features may indicate dynamical shaping by forming planets.
\vspace{6pt}

In recent years, infrared and H$\alpha$ differential imaging have led to the discovery of candidate protoplanets (planets assembling in their natal disks), such as LkCa15 b and PDS70 b, which have wide ($\gtrsim15$ AU) orbital separations and low contrast
\citep[$\sim10^{-2}$-$10^{-3}$; e.g][]{2012ApJ...745....5K,2015Natur.527..342S,2018A&A...617A..44K,2018ApJ...863L...8W}.
These detections represent just the tip of the iceberg of expected protoplanet properties.
While protoplanets are expected to have relatively low contrasts \citep[$\sim10^{-3}$-$10^{-7}$; e.g.][]{2015ApJ...803L...4E,2015ApJ...799...16Z} compared to mature ($\sim$ Gyr) planets, current high contrast imagers\footnote{e.g. HST \citep[][]{2003AJ....125.1467S},
Keck/NIRC2 \citep{2019AJ....157...33M}, VLT/SPHERE \citep{2019arXiv190204080B}, Gemini/GPI \citep{2008SPIE.7015E..18M}, Subaru/SCExAO \citep{2015PASP..127..890J}, Magellan/MagAO \citep{Close:2012}} only probe the bright end of this range \citep[$\sim10^{-4}$ contrast at a few $\lambda/D$; e.g.][]{2018AJ....156..156X,2015A&A...576A.121M}.
Furthermore, due to the large distances to nearby star forming regions ($\sim140$ pc), 8-10 meter class telescopes only resolve spatial separations of tens of AU in the infrared where protoplanets are expected to be bright.
We are thus currently limited to wide-separation detections of only high-luminosity planets, and we cannot resolve the regions of protoplanetary disks where giant planet formation is thought to be most efficient (down to $\sim$3 AU).

\vspace{6pt}

While sophisticated models have been developed to relate both old ($\sim$Gyr) and young ($\sim$Myr) planet properties to their formation histories \citep[e.g.][]{2016ApJ...832...41M,2017A&A...608A..72M,2012ApJ...745..174S}, and to predict how circumplanetary disk properties vary with the planet formation environment \citep[e.g.][]{2017MNRAS.464.3158S}, we lack the data to nail down their details.
Our limited observational capabilities leave a number of unanswered fundamental questions regarding planet formation and disk-planet interactions:
 
\vspace{6pt}
\begin{itemize}[leftmargin=*]
    \item \textbf{How are gas-giant planets assembled?} What is their distribution of accretion rates and entropies? How steady or stochastic is planetary accretion? What is the infall geometry?\vspace{1pt}
    \item \textbf{What are early planet formation conditions?} What is the distribution of protoplanet compositions, temperatures, and surface gravities? What is their distribution of metallicites and abundance ratios?\vspace{1pt}
    \item \textbf{How do protoplanets interact with their disks and with each other?} What is the relationship between protoplanet, circumplanetary disk, and protoplanetary disk properties? Are the gaps and rings seen in protoplanetary disks caused by young planets? 
\end{itemize}
\vspace{6pt}

\textbf{Answering these questions requires high contrast imaging and spectroscopy of protoplanets and protoplanetary disks at $\mathbf{\sim3-50}$ AU separations.}

\section{Observing Giant Planet Formation with Imaging and Spectroscopy}
\vspace{-8pt}
Building a census of well-characterized protoplanets with orbital separations down to $\sim$3 AU, and imaging protoplanetary disks on the same spatial scales will achieve the following science goals: 
\vspace{6pt}
\begin{enumerate}[leftmargin=*]
    \item \textbf{Determine circumplanetary accretion mechanisms} by mapping the distribution of protoplanet luminosities, measuring variability, and resolving accretion-tracing emission lines\vspace{1pt} 
    \item \textbf{Establish early formation properties} by measuring protoplanet compositions, temperatures, and surface gravities\vspace{1pt}
    \item \textbf{Study disk-planet interactions directly} by characterizing protoplanetary disk structures and connecting them to planet detections or meaningful upper limits
\end{enumerate}
\vspace{6pt}

\begin{figure}
{\caption{Science goals and technical requirements for characterizing planet formation directly\vspace{-10pt}}\label{tab:reqs}}
{\includegraphics[width=6.5in]{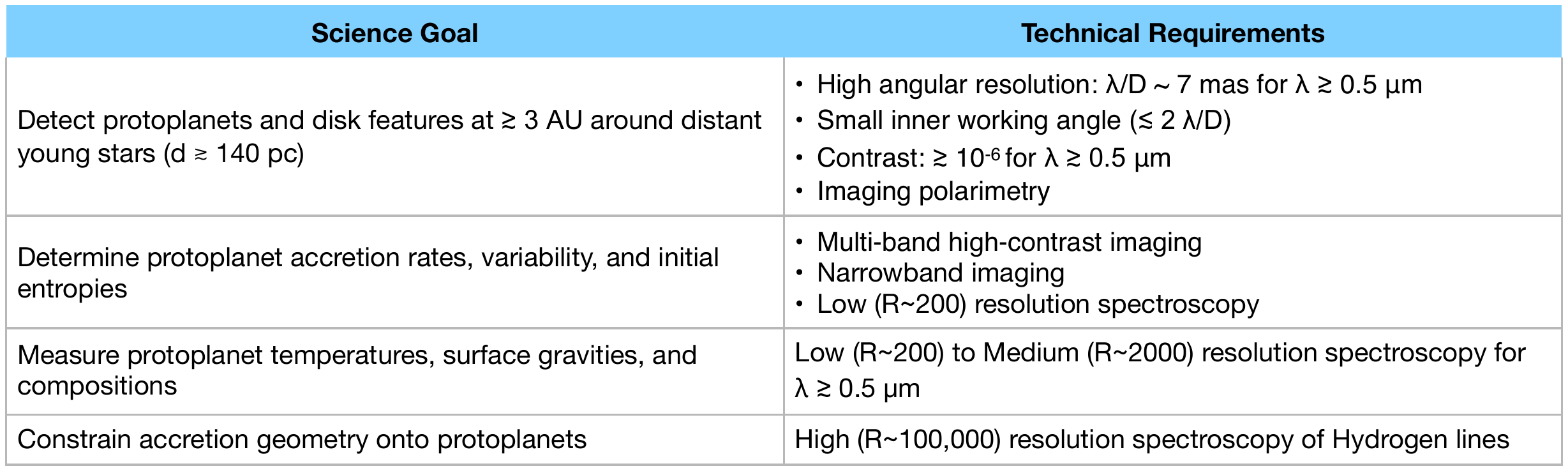}\vspace*{-10pt}}
\end{figure}

This work requires high resolution and contrast, and a range of spectral resolutions from the visible through the mid-infrared (Figure \ref{tab:reqs}), requirements that we will meet in the coming decade.
Upgraded adaptive optics systems and instruments on 8-10 meter telescopes\footnote{e.g. Gemini/GPI 2.0 \citep{2018SPIE10702E..44C}, Keck/KAPA \citep{2016SPIE.9909E..15W}, Keck/KPIC \citep{2018SPIE10703E..06M}, Magellan/MagAO-X \citep{2018SPIE10703E..09M}, Keck/SCALES \citep{2018SPIE10702E..A5S}, Subaru/SCExAO \citep{2018SPIE10703E..50S}} will improve our achievable contrast at wide ($\sim10$ AU) orbital separations, accessing lower planet luminosties than we can currently probe.
Space-based imagers\footnote{e.g. JWST NIRCam and NIRISS in the infrared, and WFIRST CGI at H$\alpha$} will push contrast even higher for wider  ($\sim$15 AU) separations.
These data will guide first-light observations on TMT and GMT.
GMT/TMT planned instruments\footnote{e.g. GMT: GMagAO-X, GMTIFS, GMTNIRS, TIGER; and TMT: IRIS, MICHI, MODHIS, NIRES, PSI} will transform protoplanet detections, providing fine angular resolution, tight inner working angle, deep contrast, and wide spectral coverage.

\vspace{6pt}

\begin{figure}[h]
\caption{Left: Predicted contrasts for 1 $\mathrm{M_J}$ planets with accretion rates of $10^{-7}$-$10^{-5}~\mathrm{M_J~yr^{-1}}$ (blue circles and red squares) as well as high initial entropy (``hot-start"; yellow triangles) and low initial entropy (``cold-start"; green diamonds) 1 Myr old photospheres around a 2 Myr solar analog (K5V spectral type).
Horizontal lines indicate estimated contrasts for Keck/KAPA and TMT/PSI at $\sim1$-$2~\lambda/D$. 
Right: Predicted H$\alpha$ contrast (colorscale) as a function of accretion rate ($\mathrm{\dot{M}}$) and spectral resolution for a $1~\mathrm{M_J}$, $\mathrm{R_p}=1.6~\mathrm{R_J}$ protoplanet, with a circumplanetary disk inner radius of $2~\mathrm{R_p}$. Dotted lines indicate factors of 10 in contrast.
Stars show observed H$\alpha$ contrasts for LkCa 15 b and PDS 70 b at MagAO's spectral resolution (corresponding to $\mathrm{\dot{M}}$ values for the assumed 2 Myr old solar analog / K5V spectral type and $1~\mathrm{M_J}$ planet mass). The solid line shows MagAO's achievable contrast at 100-200 mas ($\sim$15-30 AU at 140 pc). MagAO-X and GMagAO-X projected contrasts lie off the bottom of the plot ($10^{-5}$-$10^{-6}$ at 100 mas and $10^{-7}$-$10^{-8}$ at 15-30 mas, respectively).\vspace{-0pt}}\label{fig:predcont}
{\includegraphics[width=6.5in]{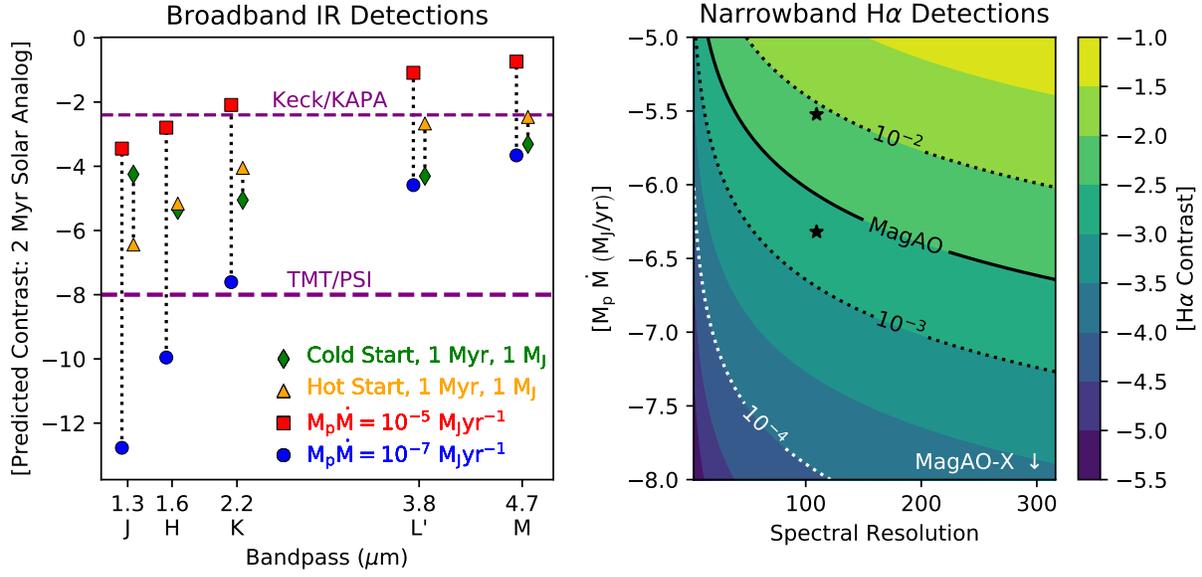}\vspace{-25pt}}
\end{figure}

These facilities will discover lower planet masses / accretion rates than current facilities (Figure \ref{fig:predcont}), resolve $\sim$3 AU orbital separations, characterize photospheres and accretion geometries (Figure \ref{fig:lines}), and image the efficient planet-forming regions of circumstellar disks for the first time (Figure \ref{fig:diskim}).
We describe this scientific progress in more detail in the following subsections.


\vspace{-10pt}
\subsection{Circumplanetary Accretion}
\vspace{-6pt}
Protoplanet emission may be a combination of the protoplanetary photosphere, the circumplanetary disk, and accretion shocks, with the relative proportions depending on the specific configuration and evolutionary state of the system.
Spectral energy distributions and spectra will disentangle these multiple components and characterize them in detail.

\vspace{6pt}
Current observations of circumplanetary accretion come from a small number of protoplanet detections with large orbital separations and high accretion luminosities \citep[e.g.][]{2012ApJ...745....5K,2015Natur.527..342S,2018ApJ...863L...8W}, as well as non-detections which place interesting constraints on the product of their mass and accretion rate \citep[$\mathrm{M_p\dot{M}}$; e.g.][]{2017AJ....154...73R}.
Imaging and spectroscopy from the visible through the mid-infrared will expand this sample over the next decade, reaching $\mathrm{M_p\dot{M}}\lesssim10^{-7}~\mathrm{M_J^2~yr^{-1}}$ and placing meaningful constraints on circumplanetary accretion mechanisms.

\vspace{6pt}
Protoplanet spectra trace their underlying accretion physics (e.g. accretion rate, shocks, circumplanetary disk properties).
Observing the infrared and Hydrogen line fluxes of many protoplanets (Figure \ref{fig:predcont}) will constrain the distribution of accretion luminosities (and $\mathrm{M_p\dot{M}}$'s), measuring the timescale of planet formation directly.
Imaging and spectra will also characterize circumplanetary disks (e.g. temperature, size, inner radius), which are predicted to have a wide range of properties depending on the circumstellar disks in which they reside \citep[e.g.][]{2015ApJ...801...84G,2017MNRAS.464.3158S}.
These data may reveal dependencies in accretion on parameters such as orbital semimajor axis, protoplanetary disk structure, and stellar metallicity, which can be compared to theoretical predictions to test formation scenarios.

\vspace{6pt}

The physics of accretion shocks affect recently formed planets' initial entropies and luminosity evolution \citep[e.g.][]{2012ApJ...745..174S,2008ApJ...683.1104F}. 
Characterizing protoplanets that have accreted most of their mass (whose emission is largely photospheric, rather than circumplanetary) using imaging and spectroscopy (R$\sim200-2000$) will yield bolometric luminosities, testing this directly.
Measuring the population of initial entropies will constrain trends in energy dissipation during formation, informing our understanding of accretion shocks.
\vspace{6pt}

\begin{figure}
\floatbox[{\capbeside\thisfloatsetup{capbesideposition={left,center},capbesidewidth=3.75in}}]{figure}[\FBwidth]
{\caption{
Example H$\alpha$ line profile shapes for planets with masses of $1~\mathrm{M_J}$, radii of $\mathrm{R_p}=1.6~\mathrm{R_J}$, accretion rates of $10^{-6}~\mathrm{M_J~yr^{-1}}$, and inner magnetosphere and circumplanetary disk radii of $1.5~\mathrm{R_p}$. The line styles indicate outer magnetosphere radii of 3 $\mathrm{R_p}$ (solid line) and 6 $\mathrm{R_p}$ (dashed line). These were generated using the TORUS radiative transfer code \citep{2014ascl.soft04006H} and a Hartmann-style magnetospheric temperature structure \citep[e.g.][]{1994ApJ...426..669H}. The lines have been scaled by their peak fluxes to easily compare their shapes.} \label{fig:lines}}
{\includegraphics[width=2.75in]{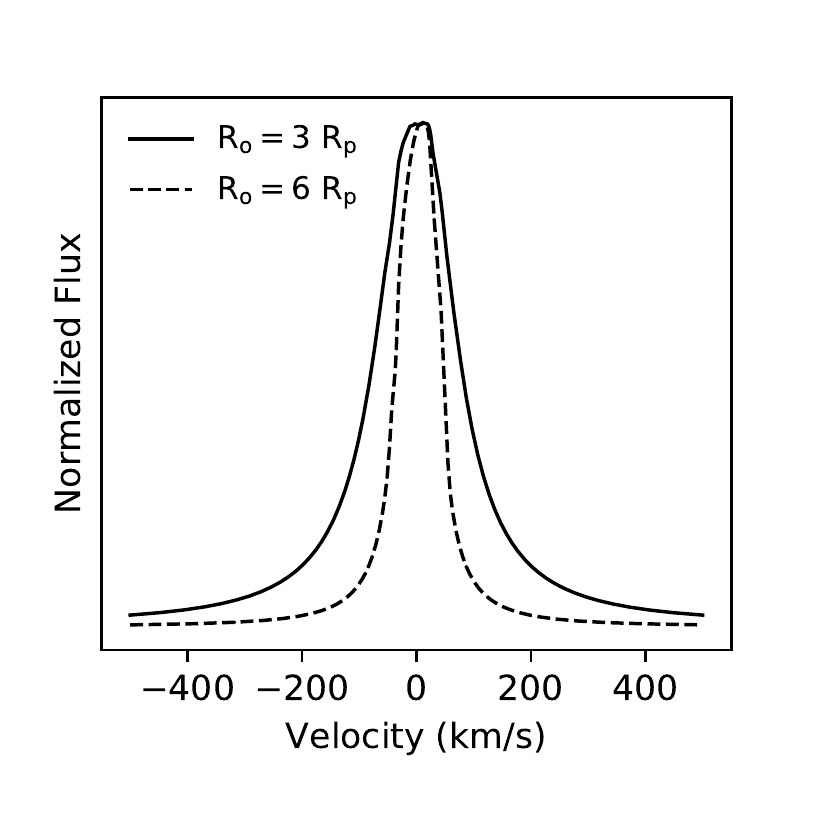}}
\vspace{-26pt}
\end{figure}

Case studies of individual protoplanets will probe infall geometry and accretion variability.
Spherical, circumplanetary disk boundary layer, and magnetospheric accretion mechanisms will have different line profiles (Figure \ref{fig:lines}).
Planned instruments for TMT and GMT\footnote{e.g. TMT: MODHIS and PSI; GMT: GMagAO-X and G-CLEF} will provide the spectral resolution (R$\sim100,000$) required to resolve these line profiles and distinguish between accretion geometries.
Time monitoring of protoplanets will also probe the detailed physics of accretion, since changes in circumplanetary disk properties \citep[e.g. viscosity; ][]{2012A&A...548A.116R} could lead to variable luminosity.

\vspace{-12pt}
\subsection{Protoplanet Photospheres and Compositions}
\vspace{-6pt}
For protoplanets dominated by photospheric emission, low- to medium-resolution near-infrared spectroscopy will determine spectral types and surface gravities \citep{2013ApJ...772...79A} and reveal any residual accretion \citep{2017AJ....154..165B}.
Atmospheric modeling and/or retrieval will constrain temperatures, surface gravities, and compositions \citep[e.g.][]{2015ApJ...804...61B,2017AJ....154...91L}, revealing early formation conditions.
Constraining compositions will be especially important for measuring bulk metallicities and elemental abundances (C/O), testing formation models and the role of ice lines \citep[e.g.][]{2016ApJ...832...41M,2018ApJ...865...32H}.
These measurements can be compared to those of intermediate age ($\sim$100 Myr) and old ($\sim$ Gyr) planets, elucidating atmospheric evolution (see white paper prepared by Bowler \& Sallum et al.).

\vspace{-12pt}
\subsection{Protoplanetary Disk Structure and Disk-Planet Interactions}
\vspace{-6pt}
As planets form they gravitationally interact with protoplanetary disks to produce density structures \citep[e.g.][]{2012ARA&A..50..211K}. 
Resolved disk observations have recently identified gaps \citep[e.g.][]{2015ApJ...808L...3A,2016ApJ...820L..40A}, spiral arms \citep[e.g.][]{2012ApJ...748L..22M,2016Sci...353.1519P,2017A&A...597A..42B}, and other asymmetries \citep[e.g.][]{2013Sci...340.1199V,2015ApJ...814L..27C,2017AJ....153..264F,2019MNRAS.485..739M} in disks.
Simulations show that these features may be produced by embedded protoplanets \citep[e.g.][]{2015ApJ...809L...5D,2015ApJ...809...93D}, and their morphologies can constrain planet properties \citep[e.g. masses and orbits;][]{2015ApJ...815L..21F}.
Multi-wavelength characterization in particular can inform planet mass estimates, as planets may preferentially keep large ($\lesssim 10~\mathbf{\mu}$m) grains from entering disk clearings  \citep[e.g.][]{2006MNRAS.373.1619R}.
\vspace{6pt}

\begin{figure}
{\caption{3D hydrodynamics and radiative transfer simulations for TMT \citep{2017ApJ...835...38D,2017ApJ...835..146D}, showing patterns in H-band scattered light produced by forming solar system giant planets on their current orbits in protoplanetary disks at 140 pc. Spiral density waves produced by Neptune (left), and gaps produced by Saturn and Jupiter (right), are clearly visible.
The images have no added noise, and assume diffraction-limited angular resolution (0.01 arcsec at H-band) and an inner working angle (IWA; the black disk at the center) twice the angular resolution. 
\vspace{-10pt}}\label{fig:diskim}}
{\includegraphics[width=4in]{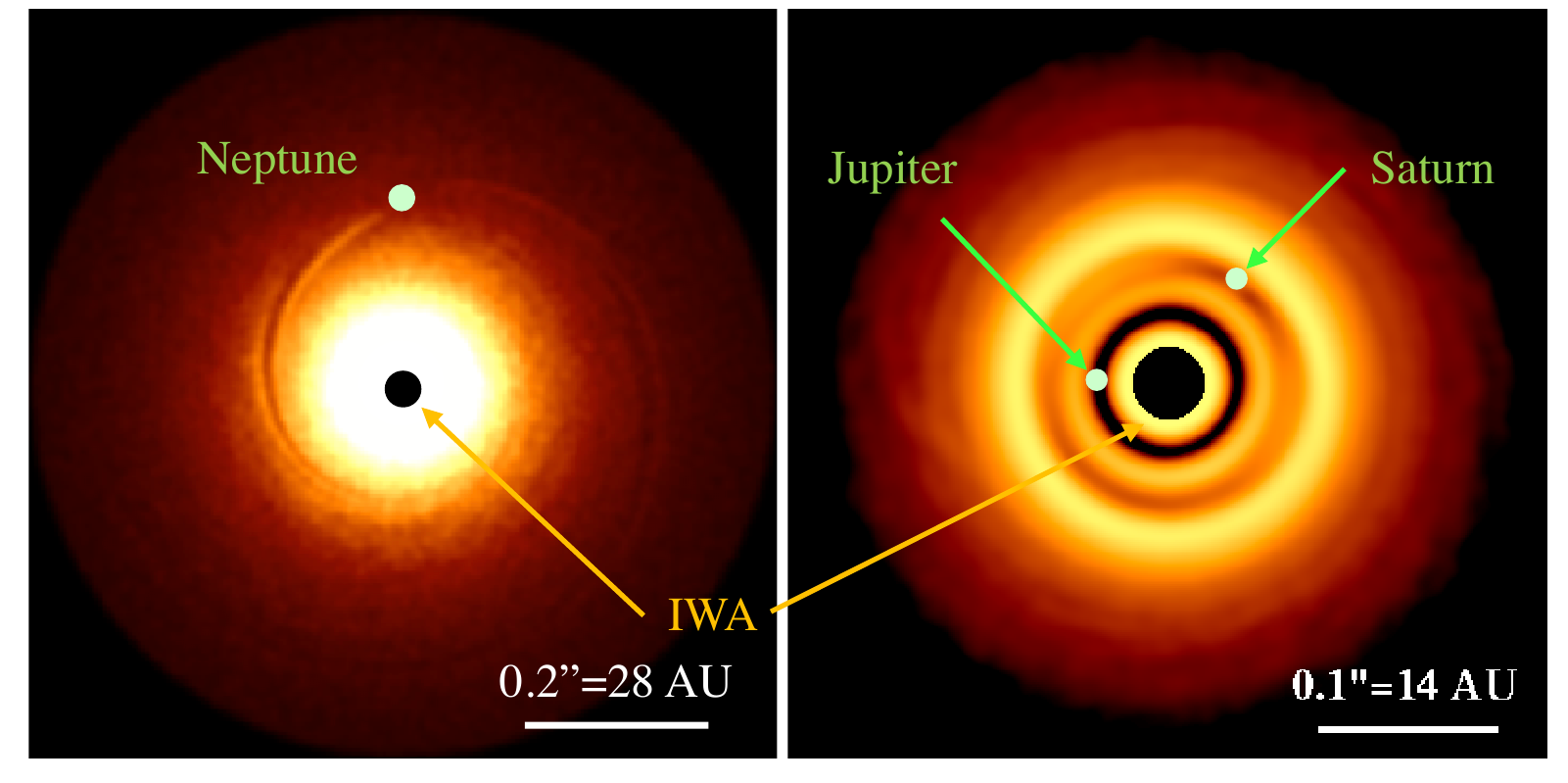}}
\end{figure}

The tight inner working angle, high contrast, and fine angular resolution of TMT/GMT will easily resolve perturbations caused by Neptune- to Jupiter-mass planets at $\gtrsim$~3 AU separations (see Figure \ref{fig:diskim}).
Polarimetric imaging will trace the scattered light, enhancing contrast and, along with multi-wavelength imaging, disentangling disk from protoplanet signals.
Observing planets with their induced disk structures will teach us the physics of disk-planet interactions \citep[e.g.][]{2019arXiv190207191M}.
Ruling out any potential planets will encourage the investigation of alternative disk sculpting scenarios, which will help us understand protoplanetary disk evolution \citep[e.g.][]{2018ApJ...862..103D}.
The white paper prepared by Jang-Condell et al. includes a more detailed discussion of disk-planet interactions and protoplanetary disk characterization.

\vspace{-15pt}
\section{Summary and Outlook}
\vspace{-10pt}
Characterizing the protoplanet and protoplanetary disk population at spatial separations down to $\sim3$ AU will enable us to finally address several fundamental questions in the field.
Measuring the distribution of protoplanet luminosities, variability, and line profiles will reveal circumplanetary accretion in detail for the first time.
Spectroscopy and polarimetry will disentangle photospheric emission from circumplanetary disk and scattered light signals, and will constrain atmospheric compositions that can be used to test formation scenarios.
Connecting these planet detections with protoplanetary disk features will directly probe planet-disk interactions.
This work will be possible thanks to instrumentation advances over the next decade, beginning with upgrades to 8-10 meter adaptive optics systems and instruments, continuing with high contrast space-based observatories, and culminating in the 30-meter telescopes.
Together these facilities will give us the most detailed and comprehensive view of the physics of giant planet formation.

\pagebreak
\let\oldbibliography\thebibliography
\renewcommand{\thebibliography}[1]{\oldbibliography{#1}
\setlength{\itemsep}{2pt}}
\bibliographystyle{aasjournal}
\bibliography{wp}

\end{document}